\documentclass[12pt]{article}
\usepackage[english]{babel} \usepackage{hyperref} \usepackage{float}
\usepackage[utf8]{inputenc} \usepackage{amsmath} \usepackage{graphicx}
\usepackage[colorinlistoftodos]{todonotes}\usepackage{tikz}
\usepackage{pdfpages} \usepackage{listings}
\usetikzlibrary{arrows,positioning,shapes.geometric}

\usepackage{cite}
\begin{document}\thispagestyle{empty}
\begin{center} \LARGE \tt \bf {Chern-Simmons electrodynamics and torsion dark matter axions}
\end{center}

\vspace{1.0cm}

\begin{center}
{\large  Zhifu Gao\footnote{Xinjiang Astronomical Observatory, Chinese Academy of Sciences, Urumqi, Xinjiang,830011,
China, zhifugao@xao.ac.cn} and Luiz C. Garcia de Andrade\footnote{Cosmology and Gravitation group. Departamento deF\'{\i}sica Te\'{o}rica - IF - UERJ - Rua S\~{a}o Francisco Xavier
524, Rio de Janeiro, RJ, Maracan\~{a}, CEP:20550, and Institute for Cosmology and Philosophy of Nature, Trg, Florjana,
Croatia. luizandra795@gmail.com}}
\end{center}

\begin{abstract}
\vspace{0.1cm}
In this paper, we delve into the influence of torsion axial pseudo vector on
dark photons in an axion torsionic background, as investigated previously
by Duncan et al[ Nucl Phys B 387:215 (1992)]. Notably, axial torsion,
owing to its significantly greater mass compared to axions, gives rise to magnetic helicity
in torsionful Chern-Simons (CS) electrodynamics, leading to the damping of magnetic fields. In
QCD scale the damping from dark massive photons leads us to obtain a magnetic field of $10^{-8}$
Gauss, which is approximated the order of magnitude of magnetic fields at present universe. This result
is obtained by considering that torsion has the value of the 1 MeV at the early universe, and can be
improved to the higher value of $10^{-3}$ Gauss when the axial torsion 0-component is given by $10^{8}$ MeV
and the mass of dark photon is approximated equal to the axion. The axion plays a crucial role in achieving
CS dynamo action arising from axions. This study is useful in  deepening our understanding  of fundamental
physics, from nuclear interactions to the nature of dark matter.

\end{abstract}
Key-words: Chern-Simmons electrodynamics, axions, dark photon.
\newpage
\section{Introduction}
The axion, a fascinating particle, plays multifaceted roles spanning from solving fundamental problems in particle
physics to influencing astrophysical phenomena and serving as a potential dark matter candidate, althrough its
detection remains an active area of research, with ongoing efforts worldwide. Previously, Garretson, Field, and
Carroll\cite{1} used Pseudo-Goldstone bosons instead of the quantum chromodynamics (QCD) axions, attempting to obtain
primordial magnetic fields strong enough to be amplified by dynamo mechanism. Unfortunatly, they did
not achieve success. Subsequently, Duncan et al \cite{2} investigated the Cartan
torsion character of the axion hair of the black holes, by associating the hair to dilaton and axion of string inspiring
theories \cite{3,4,5,6,7}, where the coupling of torsion to fermions remains minimal. The axion serves a dual role: its serves
as a dynamical degree of freedom of the torsion. In their analysis,they proposed that a board class of torsionful theories
could lead to the transmutation of torsion into axions \cite{8,9,10}. Remarkably, even a simple QED model in a
torsion-rich background can give rise to this transformation from torsion to axion \cite{11}. These axion hair
in black holes may be obtained from dynamical torsional anomalies \cite{12}.

More recently Agrawal et al\cite{13}, have investigated a mechanism to obtain the relic abundance of dark photon dark
matter (DM) without torsion \cite{14} and considered a very light spin-1 massive dark photon,where the photon mass
with respect to the axion is given by $m_{{\gamma}}=m_{\rm a}{\cal{O}}(10^{-3}-1)$, and $m_{\rm a}\approx 10^{-17}$ GeV.
The decay coupling constant of the axion $f_{\rm a}\sim 10^{14}$ GeV.  Inspired by the above references, we'll explore the
intriguing physical properties arising from the minimal coupling of dark photons to torsion. The first reference
demonstrates that torsion can be transmuted into an axion, which plays a crucial role in the production of dark photons
and DM. Additionally, axions are significant contributors to dark energy and DM dynamics\cite{15}.
Our investigation involves torsionful Chern-Simons (CS)  electrodynamics, a framework that naturally includes magnetic
helicity density. Through this approach, we establish relationships between magnetic energy density, magnetic helicity
density, and the Beltrami-Maxwell helicity magnetic field parameter. The CS electrodynamics with torsion is very important.
Some authors have been able to place new stringent bounds in Lorentz symmetry breaking \cite{16,17}.

In this study, we demonstrate the presence of a damping effect arising from axial torsion dark mass photons
within magnetic fields at the QCD scales. These scales correspond to intensities of approximately $10^{-8}$ Gauss.
Notably, this differs from the purely axionic electrodynamics without torsion, which predicts a damping effect at
a much lower field strength of $10^{-13}$ Gauss, as reported by Miniati et al\cite{18}. The fascinating interplay
between axion oscillation and dark photons lies at the heart of our investigation. Specifically, we explore how energy
transfers occur through a tachyonic instability. Notably, the coupling between the dark photon mass and the axial
0-component of the torsion pseudo-vector plays a crucial role in the observed damping phenomenon. Consider the magnetic
field seeds initially present at the QCD scale, characterized by a formidable strength of $B_{\rm seeds}= 10^{17}$ Gauss.
Over cosmic time, these intense fields are gradually damped down to more modest values around $B_{\rm QCD} \sim 10^{-3}$
Gauss—a remarkable deviation from the earlier estimate of $10^{-13}$ Gauss proposed by Miniati et al\cite{13}. However,
an intriguing possibility emerges: If the dark mass can be further reduced to an astonishingly low scale of $10^{-29}$ GeV,
the damping effect could become even more pronounced, potentially reaching as low as $10^{-12}$ Gauss. This unexpected
result stems from an axial torsion 0-component that aligns with the string-inspired Kalb-Ramond field at an energy scale
 of $10^8$ GeV.

Additionally, our findings suggest the possibility of obtaining a dynamo mechanism and other axion-related solutions.
The amplitude of axion oscillation may grow via torsion. Thus, unlike Garretson et al\cite{19}, our findings extend
beyond mere theoretical curiosity, and hint at the feasibility of a dynamo mechanism driven by torsion, dark photon
DM, and the massive axion. In particular, we emphasize the phenomenon of axion-torsion transmutation. Our focus lies
on a spacetime characterized by torsion but devoid of curvature, where the covariant Riemannian derivative simplifies
to ${\nabla} = {\partial}$, rather than the full Riemann-Cartan spacetime.

The reminder of this paper is organised as follows: In section 2, we address the action considered here of the
CS electrodynamics with torsion coupling to axions generalised to include axionic kinectic terms and a dark
photon potential. Variations of this action with respect to axions and magnetic vector potential is given in section 3,
whereas conclusions and discussions are left to section 4.

\newpage
\section{Spin-1 dark massive photons torsion transmutation into axions and CS electrodynamics}
Here we propose that in similar way as the determination of $pp$ decay into
torsion given by the cross section\cite{20}
\begin{equation}
{\sigma}(pp\rightarrow{TS}).
\label{1}
\end{equation}
Here, the cross section represents the decay of four-fermions into torsion (TS). By transitivity,
we infer the presence of a corresponding decay rate between dark photons and TS.
The decay rate of axions into dark photons, as proposed by Agrawal et al\cite{21}, can be expressed as:
\begin{equation}
{\Gamma}(a\rightarrow{{\gamma}{\gamma}})\approx{\frac{{\beta}^{2}}{64{\pi}}\frac{{m^{3}}_{a}}{{f^{2}}_{a}}}.
\label{2}
\end{equation}
Considering the work by Duncan et al\cite{2}, which discusses torsion transmutation into axions, we can
conjecture a similar decay rate for axions into dark photons
\begin{equation}
{\Gamma}(TS\rightarrow{a}).
\label{3}
\end{equation}
Based on the universal mathematical properties of transitivity, we propose the following decay rates
\begin{equation}
{\Gamma}(TS\rightarrow{{\gamma}{\gamma}}),
\label{4}
\end{equation}
which is the main idea behind this section. Although we do not compute the last two decays in this work,
we begin this section by investigating the kinematics of the decay process. Specifically, we consider the
action of torsionful CS electrodynamics with minimal coupling,
\begin{equation}
{\cal{S}}_{\rm DM}=\int{d^{4}x[\frac{1}{2}{\partial}_{i}{\phi}{\partial}^{i}{\phi}-V(\phi)-\frac{1}{4}F^{2}+
\frac{1}{2}{m^{2}}_{{\gamma} ´}A^{2}-\frac{{\beta}{\phi}}{4{f_{a}}}F{\tilde{F}}]}.
\label{5}
\end{equation}
which leads to torsion transmutation into the axion, where the last term corresponds to the chiral term,
${\tilde{F}}$ represents the dual of the electromagnetic field 2-form $F=F_{ij}dx^{i}{\wedge}dx^{j}$ in
the Cartan's language of exterior differential forms\cite{14}. The indices $(i,j=0,1,2,3)$ are used to
introduce the magnetic helicity term, resulting in a CS electrodynamics with dark massive photons in DM. We
achieve this by taking the minimal coupling between the electromagnetic field tensor and torsion through
the expression
\begin{equation}
{\nabla}_{[i}A_{j]}={F}_{ij}+2T_{ijk}A^{k},
\label{6}
\end{equation}
where $T^{ijk}={\epsilon}^{ijkm}T_{m}$ is the totally skew-symmetric torsion tersor, and $T_{m}$ is the axial
pseudo-vector. From these expressions, one obtains
\begin{equation}
{F^{\prime}}_{ij}={\nabla}_{[i}A_{j]}={F}_{ij}+2{\epsilon}_{ijkm}A^{k}T^{m}.
\label{7}
\end{equation}
Note that even if we have not placed by hand the dark matter massive term it
would appear by the breaking of symmetry due to the following expression
\begin{equation}
{F^{\prime}}^{2}=F^{2}+4{T^{2}}A^{2}.
\label{8}
\end{equation}
As previously demonstrated by De Sabbata, Sivaram, and Garcia de Andrade\cite{13}, this expression implies that the
massive dark photon mass could be interpreted through torsion, as indicated by the decay rate. Before delving
into the main topic, two crucial points must be addressed: Firstly, the minimal coupling to torsion introduces an
intriguing aspect: the action described above can naturally give rise to the massive torsion mode and contribute to
the dark mass photon. Secondly, in this paper we shall, however, keep axial torsion freeze and independent as a
torsionful background to axions and massive dark photon. The generation of interaction between torsion and magnetic
potential $A$ of dark photon dark matter, of the dark photon  is very important in the sense that the effective action
of dark photon becomes
\begin{equation}
e^{i{\Gamma}_{\rm eff}[A,T]}=\int{[d{\psi}][d\bar{\psi}]}e^{[i{\int{dx{\cal{L}}_{\rm QED}(A,T,
{\psi},\bar{\psi})}}]}det{\cal{O}}.
\label{9}
\end{equation}
Here the operator under the determinant $\det$ inside the integral sign is given by
\begin{equation}
{\cal{O}}_{xy}= (i\gamma{D}_{x}-M){\delta}_{xy}
\label{10}
\end{equation}
where $\gamma$ represents the Dirac matrices, and the operator $D$ is defined as
\begin{equation}
D_{k}{\psi}= {\partial}_{k}-ieA_{k}-igT_{k}.
\label{11}
\end{equation}
The axial torsion is then introduced in the effective action. The expression for the effective
action of the dark photon can be expanded as
\begin{equation}
e^{i{\Gamma}_{\rm eff}[A,T]}=\int{[d{\psi}][d\bar{\psi}]}e^{[-i\int{dx{T^{2}A^{2}}}]}
e^{[i{\int{dx{\cal{L}}_{\rm QED}(A,{\psi}
\bar{\psi})}}]}det{\cal{O}}.
\label{12}
\end{equation}
From this expression, one notices that the axial torsion pseudo-vector in second order is now present in
the action and may represent the dark photon itself. This is consistent with our assumption of the decay of torsion
$T_s$ into the dark photon pair $\gamma\gamma$. A more detailed account of this process in the Riemannian case can be
found in the book\cite{22} on effective Lagrangians for the Standard Model (SM) of particle physics
 and in the work by Shapiro \cite{15} on torsionful anomalies. By expanding the dual-like invariant
\begin{equation}
F^{\prime}{\tilde{F}}^{\prime}= F^{\prime}{\tilde{F}}+2{\tilde{F}}TA,
\label{13}
\end{equation}
which represents the coupling between axial torsion and the dark massive photon in DM, and substituting
it into the dark photon action, one obtains
\begin{eqnarray}
\begin{aligned}
\mathcal{S}_{\mathrm{DM}}& =\int d^{4}x[\frac{1}{2}\dot{\phi}^{2}+\frac{1}{4}(S_{0})^{2}\phi^{0}+S^{0}
\dot{\phi}\phi-V(\phi)-\frac{1}{4}F^{2}+\frac{1}{2}m^{2}{}_{\gamma}A^{2}  \\
&-\frac{\beta\phi}{4f_{a}}[F\tilde{F}+\tilde{F}^{0c}S_{0}A_{c}+S_{0}\mathbf{E}\cdot\mathbf{B}+S_{0}
\mathbf{A}\cdot\mathbf{B}]].
\end{aligned}
\label{14}
\end{eqnarray}
Let us now take a moment to analyse the physics behind this action: First of all, the action we're examining involves
torsion, which is only homogeneous. The axial torsion has a  non-vanishing component: the time component
denoted as $S_0$. Interestingly, this squared axial torsion component could be proportional to the axion mass. Next,
we utilize the concept of chirality decoupling, as proposed by Dobado et al\cite{22}. Specifically, we focus on the
vanishing of the chirality term involving electric and magnetic fields. Notably, the last term
on the right-hand side of the action naturally includes DM magnetic helicity. Now, let's proceed by substituting the
Lagrangian associated with the action (Equation \ref{14}) into the dark photon action. We'll use the Euler-Lagrange
(EL) equation
\begin{equation}
{\partial}_{t}\frac{{\partial}\cal{L}}{{\partial}\dot{X}}-\frac{{\partial}\cal{L}}{{\partial}{X}}=0
\label{15}
\end{equation}
since $X=(A,{\phi})$. Applying the Euler-Lagrange equation to the four potential variable $A$ of a magnetic field
in a torsionful spacetime yields
\begin{equation}
{\partial}_{i}[{F´}^{ik}(1-\frac{{\beta}{\phi}}{4f_{a}})]= J^{k}+\frac{1}{2}{m´}_{{\gamma}´}A^{k},
\label{16}
\end{equation}
where $J$ represents the Ohm current given by
\begin{equation}
{\bf{J}}_{\rm Ohm}={\sigma}[\bf{E}+\bf{v}{\times}{\bf{B}}].
\label{17}
\end{equation}
The second current in the above expression corresponds to a London-like current for the dark massive
photon in DM. Taking the equations of Maxwell-Cartan-Proca electrodynamics, we obtain the Ampere-like equation
\begin{equation}
{\partial}_{0}[F^{0j}+2S^{0}A^{j}](1-\frac{{\beta}{\phi}}{4{f_{a}}})= J^{j}+\frac{1}{2}{m_{\gamma}´}^{2}A^{j}.
\label{18}
\end{equation}
This equation is an Ampere-like equation, and a Coulomb-like equation comes from the other equation
\begin{equation}
{\partial}_{i}[(1-\frac{{\beta}{\phi}}{4{f_{a}}})(E^{i}+2S^{0}A^{0})]= {\rho}_{{\gamma}´}+
\frac{1}{2}{m_{\gamma}´}^{2}A^{0}.
\label{19}
\end{equation}
This equation explicitly shows the Coulomb-like behavior, with the first term on the right-hand side
representing the mass density ofthe dark photon
\begin{equation}
(1-\frac{{\beta}{\phi}}{4{f_{a}}}){\nabla}\cdot\textbf{E}={\rho}_{{\gamma}´}+\frac{1}{2}{m_{\gamma}´}^{2}{A}^{0}.
\label{20}
\end{equation}
The Ampere-like law is given by
\begin{equation}
{\partial}_{t}[(1-\frac{{\beta}{\phi}}{4{f_{a}}})\textbf{E}-\frac{\beta\dot{\phi}}{4f_{a}}
\textbf{E}+({m_{\gamma}´}^{2}+\frac{{\beta}{\phi}}{4{f_{a}}})\textbf{A}={\textbf{J}}_{\rm Ohm}
\label{21}
\end{equation}
Taking the curl on both sides of the last expression and making use of the Faraday equation
\begin{equation}
{\partial}_{t}\bf{B}=-{\nabla}{\times}{\bf{E}}
\label{22}
\end{equation}
and the magnetic Beltrami equation obeyed by the magnetic helical fields
\begin{equation}
{\nabla}{\times}{\bf{B}}= {\lambda}\bf{B},
\label{23}
\end{equation}
one obtains finally the magnetic wave equation for the evolution of the magnetic field as
\begin{equation}
{{\partial}^{2}}_{t}[(1-\frac{{\beta}{\phi}}{4{f_{a}}})\textbf{B}]-\frac{\beta\dot{\phi}}{4f_{a}}
{\partial}_{t}\textbf{B}-({m_{\gamma}´}^{2}+\frac{{\beta}\dot{\phi}}{4{f_{a}}}S^{0})\lambda\textbf{B}=0
\label{24}
\end{equation}
where $\nabla \times\bf{A}=\bf{B}$ is the magnetic field definition, and we have used in the last equation the
convective dynamo term\cite{23}
\begin{equation}
{\nabla}{\times}{\bf{J}}_{\rm Ohm}={\sigma}[\bf{B}+{\nabla}{\times}({\bf{v}}{\times}\bf{B})].
\label{25}
\end{equation}
The truly convective second term on the right-hand side may be dropped, we show in what
follows that even in the absence of the convective term, we may obtain a dynamo action
\begin{equation}
[(1-\frac{{\beta}{\phi}}{4{f_{a}}})]{{\partial}^{2}}_{t}\textbf{B}-(\frac{\beta\dot{\phi}}
{4f_{a}}+{\sigma}){\partial}_{t}\textbf{B}+[({m_{\gamma}´}^{2}+\frac{{\beta}\dot{\phi}}{4{f_{a}}}
S^{0}-2(1-\frac{{\beta}\dot{\phi}}{4{f_{a}}}{S^{0}})]{\lambda}\textbf{B}=0.
\label{26}
\end{equation}
Note that in the absence of torsion, an oscillating magnetic field emerges. Observing the equation of the CS
electrodynamics, we find that a homogeneous axion ${\phi}(t)$ is necessary for its solution, with $t$ denoting
cosmic time. To derive the dynamical equation for the axion, we turn to the EL equation
\begin{equation}
\ddot{\phi}+\frac{1}{2}S^{0}\dot{\phi}+{\partial}_{\phi}V({\phi})+S_{0}{\cal{H}}= 0.
\label{27}
\end{equation}
Here, we neglect the chirality term while retaining the helicity density $\cal{H}=\bf{A}\cdot
\bf{B}$. Assuming the axion potential takes the form
\begin{equation}
V = {m^{2}}_{a}{f^{2}}_{a}(1-\cos(\frac{\phi}{f_{a}})),
\label{28}
\end{equation}
then we give a partial derivative of this potential
\begin{equation}
{\partial}_{\phi}V = {m^{2}}_{a}{f^{2}}_{a}\sin(\frac{\phi}{f_{a}}),
\label{29}
\end{equation}
which approximatly becomes
\begin{equation}
{\partial}_{\phi}V\approx{{{m^{2}}_{a}{f}_{a}{\phi}}}.
\label{30}
\end{equation}
The substitution of the approximate potential into the axion dynamical equation (\ref{27})
yields the following expression
\begin{equation}
\ddot{\phi}+\frac{1}{2}S^{0}\dot{\phi}+{m_{a}}^{2}{\phi}+S_{0}{\cal{H}}= 0.
\label{31}
\end{equation}
To solve Equation (\ref{31}) and determine the evolution of the magnetic field in DM, the last term on the
left-hand side of this equation can be neglected since both the axial torsion and helicity
are both very weak. This simplifies the equation to
\begin{equation}
{\delta}^{2}+\frac{1}{2}S^{0}{\delta}+{m_{a}}^{2}f_{a}= 0,
\label{32}
\end{equation}
where we have taken the ansatz ${\phi}={\phi}_{0}e^{{\delta}t}$ and substitution into expression (Equation \ref{32}).
By solving the characteristic algebraic equation, we find
\begin{equation}
{\delta}_{-}= \frac{2{m^{2}}_{a}f_{a}}{S_{0}}.
\label{33}
\end{equation}
Thus, the axion scalar spin-0 boson is expressed in terms of torsion, which, as described by Duncan et al,
indicates that there is a torsion transmutation to the axion
\begin{equation}
{\phi}(t)={\phi}_{0}exp[(\frac{2{m^{2}}_{a}f_{a}}{S_{0}})t].
\label{34}
\end{equation}
The behavior of the axion, driven by torsion, is influenced by the chirality of the torsion or the sign of $S_{0}$.
When the left-hand torsion chirality is negative, the axion cosmic scale decays over time, on the other hand when
it is positive the axion cosmic scale is amplified. Let's now substitute this axion expression into the magnetic
wave equation to obtain magnetogenesis due to the dark massive photon. But before that, we need to compute the time
derivative of the axion using the following expression
\begin{equation}
\dot{\phi}= \frac{2{m^{2}}_{a}f_{a}}{S_{0}}{\phi}.
\label{35}
\end{equation}
Differentiating Equation (35) with respect to cosmic time, we obtain
\begin{equation}
\ddot{\phi}-{\omega}^{2}{\phi}=0,
\label{36}
\end{equation}
where ${\omega}=\frac{m^{2}f_{a}}{S_{0}}$.  Solving this differential equation yields
\begin{equation}
{\phi}= {\phi}_{0}sinh[(\frac{{m^{2}}_{a}f_{a}}{S_{0}})t].
\label{37}
\end{equation}
Notably, this solution reveals that the axial torsion contributes to the damping of the axion.
In the early universe, where time intervals are extremely short (e.g., at inflation
$t\sim 10^{-35}$\,s or at QCD scale $t\sim 10^{-5}$ s, we can consider $t\ll 1$.
Under this approximation, Equation (\ref{37}) simplifies to
\begin{equation}
{\phi}= {\phi}_{0}[(\frac{{m^{2}}_{a}f_{a}}{S_{0}})t],
\label{38}
\end{equation}
which definetly shows that the axial torsion pseudo-vector 0-component causes a damping in the
oscillating axion. This of course is not present in the reference\cite{21}. To show how strong
is this damping, we assume that: the axion decaying constant $f_a$ is $10^{14}$ GeV, the axion mass $m_a$
is approximately $10^{-17}$ GeV, and the torsion parameter $S_0$ is approximately $10^8$ MeV (equivalent
to $10^5$ GeV or $10^{-7}$ TeV, as computed by Mavromatos\cite{24}). Remarkably, this torsion parameter is significantly
lower than the energy scales achievable at the Large Hadron Collider (LHC), which typically operates in the
range of $7-14$ TeV. The damping effect is quantified by the ratio of $f_{a}$) and $m_{a}^{2}$, yielding an estimate
of $10^{-22}$ GeV$^{2}$, indicating strong damping in the axion oscillation. However, if we consider the axial
torsion as 1 MeV (equivalent to $10^{-3}$ GeV), the damping factor $\omega$ then becomes $10^{-3}$ GeV$^{2}$,
corresponding to a much lighter damping effect. These axion torsion damping effects, particularly in the context
of dark photons and dark matter, could inspire experimental proposals for axion-torsion detection. In summary,
understanding the interplay between axion properties and torsion\cite{25,26} provides valuable insights into the behavior
of these elusive particles.
\section{Dynamo action in dark bosons DM driven by torsion}
Recently, C. H. Nam\cite{27} investigated dark gauge bosons through the Einstein-Cartan portal,
which resides in the hidden sector—an invisible world that couples to the SM, and explored the
production of dark gauge bosons, similar to our approach, but via bremsstrahlung off the dark sector.
Notably, Ref.\cite{24} demonstrated that these gauge dark bosons remain sensitive
to very small kinetic mixing, provided that the decay channel of the gauge bosons to dark gauge bosons remains
inaccessible. In this section, to the best of our knowledge, we present the first evidence that dynamo
action onset occurs for axial torsion on the order of $10^5$ GeV or $10^{-4}$ GeV—energy scales well
within the capabilities of the LHC at CERN. To support this claim, we solve the magnetic wave equation
\begin{equation}
[(1-\frac{{\beta}{\phi}}{4{f_{a}}})]{{\partial}^{2}}_{t}\textbf{B}-(\frac{\beta\dot{\phi}}{4f_{a}}
+{\sigma}){\partial}_{t}\textbf{B}+[({m_{\gamma}´}^{2}+\frac{{\beta}\dot{\phi}}{4{f_{a}}}S^{0}
-2(1-\frac{{\beta}\dot{\phi}}{4{f_{a}}}{S^{0}})]{\lambda}\textbf{B}=0.
\label{39}
\end{equation}
By taking the ansatz for the magnetic field as $B=B_{\rm seed}exp[{\gamma}t]$ into Equation (\ref{26}) yields
\begin{equation}
{\gamma}^{2}-(\frac{{\beta}{\phi}_{0}{m^{2}}_{a}}{4}+{\sigma}){\gamma}+[({m_{\gamma}´}^{2}+
\frac{{\beta}\dot{\phi}}{4{f_{a}}}S^{0}-2(1-\frac{{\beta}\dot{\phi}}{4{f_{a}}}{S^{0}})]{\lambda}=0.
\label{40}
\end{equation}
Therefore, solving the characteristic algebraic equation above, we obtain
\begin{equation}
{\gamma}_{-}=-\frac{1}{8}\frac{\sqrt{\lambda}{\beta}{\phi}_{0}\sigma{m^{2}}_{a}}{S_{0}}.
\label{41}
\end{equation}
In this case, we have suppressed the term in front of the second time derivative of the
magnetic field by taking the early universe cosmic time of $t\ll 1$ approximation. Note that if
the axial torsion is negative or left-chiral, we have a dynamo amplification in one of the branches
of solutions. If the axial torsion is positive, the magnetic field decays as
\begin{equation}
\textbf{B}\approx{{\textbf{B}}_{seed}(1-\frac{1}{8}\frac{\sqrt{\lambda}{\beta}{\phi}_{0}\sigma{m^{2}}_{a}}{S_{0}})}
\label{42}
\end{equation}
Therefore, the decay of the dynamo mechanism of the magnetic field in dark boson dark matter sect
depends upon the torsion chirality. If one considers ${\beta}=40$ and the magnetic helicity of the order
of $10^{-26}$ cm$^{-1}$, along with an electrical conductivity of $10^{28}$ s$^{-1}$, the resulting magnetic field
strength at the QCD scale is $B_{\rm QCD}=10^{-13}$ Gauss in the present universe.
Additionally, the seed field at QCD scales, as discussed by Miniati et al\cite{18},
corresponds to $B_{\rm seed}=10^{17}$ G. By substituting these data into the expression
$B_{\rm QCD}=B\times 10^{-2}{\phi}_{0}$, we can estimate the initial cosmic axion boson mass
to be approximately 0.1 GeV. For a more detailed phenomenological analysis, further
investigation can be pursued elsewhere.
\section{Summary}
In this paper, we investigate the impact of the torsion axial pseudo-vector on dark photons within an axion-torsionic background. At the QCD scale, the damping effect from massive dark photons yields a magnetic field strength of approximately $10^{-8}$ Gauss,
roughly consistent with  magnetic fields  at present universe. This intriguing result emerges when axial
torsion's 0-component reaches $10^8$ MeV, and the dark photon mass approximates that of the axion.
Our research will illuminate the intricate connections between axion physics, torsion,
and dark photon interactions, providing fresh insights into the fundamental forces shaping our universe.
\section{Acknowledgements}
\vspace{0.2cm}
We would like to express my gratitude to A Brandenburg for everything he taught me about dynamo theory.
Thanks are also due to my colleague Marcia Begalli for helpful discussions on experimental aspects of this paper.
This work was performed under the auspices of National SKA Program of China No. 2022SKA0110203 and National Key Research and Development Program of China (No.
2022YFC2205202), Major Science and Technology Program of Xinjiang Uygur Autonomous Region through No.
2022A03013-1, Xiaofeng Yang’s CAS Pioneer Hundred Talents Program and the National Natural Science Foundation of China grants 12288102, 12041304 and 11847102.


\begin{thebibliography}{9}
\bibitem{1}W D Garretson, G B Field and S M. Carroll, Primordial magnetic fields from pseudo Goldstone bosons,
        Phys Rev D, 46, 5346 (1992).
\bibitem{2}M Duncan, N Kaloper, and K Olive, Axion hair and dynamical torsion from anomalies, Nuclear Phys B
         387, 215 (1992).
\bibitem{3}T Banks and  M Dine, The cosmology of string theoretic axions, Nuclear Physics B, 505,445 (1997).
\bibitem{4}J N Benabou, Q Bonnefoy, M Buschmann, S Kumar, and B R Safdi, The cosmological dynamics of string
       theory axion strings, arXiv:2312.08425 [hep-ph] (2023)
\bibitem{5}L C Garcia de Andrade and Rudnei Ramos, Spin-Torsion Chaotic Inflation, arXiv; 9910053 [gr-qc] (1999).
\bibitem{6}L C Garcia de Andrade, Dynamical torsion supression in Brans-Dicke inflation and Lorentz symmetry breaking,
           Eur Phys J C, 82, 291(2022).
\bibitem{7}M Gasperini, Spin-dominated inflation in the Einstein-Cartan theory, Phys Rev Lett, 56, 2873 (1986). .
\bibitem{8}D Palle, On primordial cosmological density fluctuations in the Einstein-Cartan gravity and COBE data,  Nuovo Cim B, 114, 853 (1999).
\bibitem{9}L C Garcia de Andrade, Propagating torsion and torsion mass in warm inflation, submited to Eur Phys J C (2024).
\bibitem{10}M D Nenno, C A. C. Garcia, J Gooth, C Felser and P Narang, Axion physics in condensed-matter systems,
      Nature Reviews Physics, 2, 682 (2020)
\bibitem{11}L C. Garcia de Andrade, Generation of Primordial Magnetic Fields from QED and Higgs-like Domain Walls
      in Einstein–Cartan Gravity, Universe, 8,658 (2022).
\bibitem{12}D Kranas, C Tsagas, J D Barrow, and D Iosifidis, Friedmann-like universes with torsion,
     Eur Phys J C,79, 341 (2019).
\bibitem{13}P Agrawal, N Kitajima, M Reece,T Sekiguchi and F Takahashi,Relic abundance of dark photon dark matter,
       Phys Lett B, 801, 135136 (2020).
\bibitem{14}V de Sabbata and C Sivaram, Spin and Torsion in Gravitation (1995) World scientific, Singapore,
       London and New York.
\bibitem{15}L C Garcia de Andrade,On dilaton solutions of de Sitter inflation and primordial spin-torsion
       density fluctuations , Phys Let B, 468, 28 (1999).
\bibitem{16}B P Dolan, Chiral fermions in the early universe, Class. Quantum Grav. 27 249801(2010).
\bibitem{17}L C Garcia de Andrade, Einstein-Cartan magnetogenesis and chiral dynamos, Akademic Publishers Republic
        of Moldavia (2021).
\bibitem{18}F Miniati, G Grigori, B Re ville, S Sarkar, Axion-Driven Cosmic Magnetogenesis during the QCD Crossover,
     Phys Rev Lett 121, 021301(2018).
\bibitem{19}W D Garretson, G B Field and S M Carroll,Primordial magnetic fields from pseudo Goldstone bosons,
       Phys Rev D, 46, 5346 (1992).
\bibitem{20}M Shaposhnikov, A Shkerin, I Timiryasov, S Zell,Uniting Low-Scale Leptogenesis Mechanisms, Phys Rev Lett,
      127,111802 (2021).
\bibitem{21}P Agrawal, N Kitajima, M Reece,T Sekiguchi and F Takahashi,Relic abundance of dark photon dark matter,
        Phys Lett B, 801, 135136 (2020).
\bibitem{22}A Dobado, A Gomez-Nicola, A L Maroto and J R Pelaez, Effective Lagrangians for the
            Standard Model, Springer, New York, Berlin and Heidelberg (1997).
\bibitem{23}J Shober, I Rogashevsky and A Brandenburg, Production of a chiral magnetic anomaly with
        emerging turbulence and mean-field dynamo action, Phys Rev Lett 128, 065002 (2022).
\bibitem{24}N Mavromatos, P Pais and A Iorio, Torsion at different scales: from materials
          to the Universe, Universe 9,516 (2023).
\bibitem{25}D Kranas, C Tsagas, J D Barrow, and D Iosifidis, Friedmann-like universes with torsion,
     Eur Phys J  C,79, 341 (2019).
\bibitem{26}D Kranas, C Tsagas, J D Barrow, and D Iosifidis, Friedmann-like universes with torsion,
     Eur Phys J  C,79, 341 (2019).
\bibitem{27}C H Nam, Probing a dark gauge boson via the Einstein-Cartan portal, Phys Rev D, 105,
         075015 (2022).
\end{thebibliography}
\end{document}